# Ionic Organic Cage-encapsulating Phase-transferable Metal Clusters

Su-Yun Zhang,[a] Zdravko Kochovski,[c] Hui-Chun Lee,[a] Yan Lu,[c,d] Jie Zhang,[a] Jian-Ke Sun,*[a] and Jiayin Yuan*[b]

Exploration of metal clusters (MCs) adaptive to both aqueous and oil phases without disturbing their size is promising for a broad scope of applications. The state-of-the-art approach *via* ligand-binding may perturb MCs' size due to varied metal-ligand binding strength when shuttling between solvents of different polarity. Herein, we applied physical confinement of a series of small noble MCs (< 1 nm) inside ionic organic cages (I-Cages), which by means of anion exchange enables reversible transfer of MCs between aqueous and hydrophobic solutions without varying their ultrasmall size. Moreover, the MCs@I-Cage hybrid serves as a recyclable, reaction-switchable catalyst featuring high activity in liquid-phase catalysis.

## Introduction

Metal clusters (MCs) are particles of < 2 nm in size and exhibit intrinsic size-dependent properties, such as discrete electronic structures, intense photoluminescence, high catalytic activity, *etc*.[1] Synthesis of MCs capable of adapting both aqueous and organic phases is favorably pursued because of a broader scope of operational environment.[2] MCs, especially of size < 1 nm practically carry all constitutional atoms on their surface, and possess a higher specific surface energy than the exhaustively explored larger metal nanoparticles (MNPs).[3] Thus they aggregate more severely without sufficient protection in solutions.[4,5] Synthetic methods of MCs are so far dominated by surface-binding ligand approach, including amphiphilic capping agents, and the oil-water phase transfer agents by either ligand exchange or modification.[2,6] A challenge associated with the ligand approaches that relies on binding active sites of MCs is that the MC core is labile and prone to size variation due to altered metal-ligand binding strength upon ligand exchange or modification.[7] In addition, these approaches suffer from a trade-off, *i.e.* binding a high-energy, catalytically active surface by polar groups of ligands restricts accessibility of reactants to these active sites during catalysis.[8] In this regard, a stabilization mechanism of MCs majorly by physical confinement and partially by surface binding would allow MCs to maintain their high catalytic activity in different solvents with little-to-no perturbation of stabilization, thus possibly breaking the trade-off.

Organic molecular cages are an emerging class of multifunctional materials featuring a discrete pore structure.[9] They are molecularly soluble and possess a wide spectrum of properties and functions due to their designable molecular architecture and intrinsic open channels. A unique merit of cages is to accommodate guest objects, such as MCs/MNPs without blocking much of their active surface sites and to "solubilize" them in liquid-phase for catalysis.[10]

Herein, we utilized ionic organic cages to synthesize and confine a series of well-dispersed, water-oil phase adaptive noble MCs of 0.5 ~0.7 nm in size (Scheme 1). Different from MCs reported previously that have limited choices of metal species and/or solvents, our example improves compatibility between the ionic cage molecule and metal species. More significantly, the highly charged cage (+12 per cage) can trap metal anion precursor to control MC growth and avoid spontaneously aggregation by electrostatic repulsion. The MCs confined in I-Cage experience anion-controlled reversible water-oil phase transfer without size perturbation. Such MC@I-Cage hybrid was explored as recyclable/switchable catalyst[11] with high activity in a liquid-phase $H_2$ release reaction.

## Experimental

**Materials and methods.**

All chemicals were obtained from commercial sources and used without further purification. Conventional transmission electron microscopy (TEM) was performed on a JEOL 2010FS transmission electron microscope operated at 120 kV. STEM was performed on a Jeol JEM-2200FS transmission electron microscope operated at 200 kV and equipped with a high-angle annular dark-field (HAADF) STEM detector. Cryo-EM was performed on a JEOL JEM-2100 transmission electron microscope (JEOL GmbH, Eching, Germany). Cryo-EM specimens were prepared by applying a 4 μl drop of a dispersion sample to Lacey carbon-coated copper TEM grids (200 mesh, Science Services) and plunge-frozen into liquid ethane with an FEI vitrobot Mark IV set at 4°C and 95% humidity. Vitrified grids were either transferred directly to the microscope cryogenic transfer holder (Gatan 914, Gatan, Munich, Germany) or stored in liquid nitrogen. Imaging was carried out at temperatures around 90 K. The TEM was operated at an acceleration voltage of 200 kV, and a defocus of the objective lens of about 2.5 – 3 μm was used to increase the contrast. Cryo-EM micrographs were recorded at a number of magnifications with a bottom-mounted 4*4k CMOS camera (TemCam-F416, TVIPS, Gauting, Germany). The total electron dose in each micrograph was kept below 20 e−/Å$^2$. X-ray photoelectron spectroscopy (XPS) studies were performed on a ThermoFisher ESCALAB250 X-ray photoelectron spectrometer (powered at 150 W) using Al K$_α$ radiation (λ = 8.357 Å). To compensate surface charging effects, the XPS spectrum was referenced to the C 1s neutral carbon peak at 284.6 eV. The solution UV-Vis absorption measurements were recorded on a Lambda 900 spectrophotometer. $^1$H and $^{13}$C nuclear magnetic resonance ($^1$H-NMR) measurements were carried out at room temperature on a Bruker ascend-400/600 spectrometer in deuterated oxide. Dynamic light scattering (DLS) was performed on a NICOMP particle sizer (model 380 PSS, Santa Barbara, CA). Zeta potential measurement was performed on Zetasizer Nano ZS90 (Malvern). Electrospray ionisation mass spectrometry (ESI-MS) was performed on 1200 Series & HCT Basic System (Bruker). Inductively coupled plasma optical emission spectrometry (ICP-OES) was performed on Varian 700-ES (Agilent ), calibrated with standard solutions.

**Synthesis of porous cycloimine cage molecule CC3.**

The synthesis of CC3 was conducted according to a previous method.[10a] A typical procedure is as follows: $CH_2Cl_2$ (10 mL) was added slowly to solid 1,3,5-triformylbenzene (0.5 g) at room temperature. Trifluoroacetic acid (10 μL) was added directly to this solution as a catalyst for imine bond formation. Finally, a $CH_2Cl_2$ solution (10 mL) of (R,R)-1,2-diaminocyclohexane (0.5 g, 4.464 mmol) was added. The vessel of the mixture solution was capped and left to stand for one week. Crystals grew on the sides of the vessel. The crystalline product was removed by centrifugation and washed with $CH_2Cl_2/CH_3OH$ mixture (v/v, 5/95) for several times, and further dried at 100 °C overnight.

**Synthesis of reduced CC3 (RCC3).**

The synthesis of RCC3 was conducted according to a previous method.[12a,c] The imine cage CC3 (463 mg) was dissolved in a $CH_2Cl_2/CH_3OH$ mixture (v/v, 1/1, 25 mL) by stirring. When this solution became clear, $NaBH_4$ (0.5 g) was added and the reaction was stirred for 15 h at room temperature. Water (1 mL) was then added, and the solution was continuously stirred for additional 9 h. The solvent was then removed under vacuum. The resulted solid was washed by water and collected by centrifugation, and the obtained solid was dried at 80 °C under vacuum overnight.

is consistent with the ESI(+)-MS analysis result [M+H]$^+$ of 1155.5 (Figure S3).

**Synthesis of metal cluster (MC)/metal nanoparticles stabilized by different supports.**

**Synthesis of Au@I-Cage-Cl.** In a typical synthesis, 9 mL of water solution containing 15 mg of I-Cage-Cl was subsequently added to 0.5 mL of aqueous $HAuCl_4$ solution (0.5 mg Au in content). The resultant mixture solution was further homogenized after aging for a few minutes. Then, 0.5 mL of aqueous solution containing 2 mg of $NaBH_4$ was immediately added into the above solution with vigorous shaking, resulting in a well transparent dispersion of Au@I-Cage-Cl.

**Synthesis of other MCs stabilized by I-Cage-Cl.** The synthetic procedure used above to prepare Au@I-Cage-Cl was followed by using 0.5 mL of aqueous containing $K_2PdCl_4$, or $K_2PtCl_4$ (The metal in content is 0.5 mg) in place of $HAuCl_4$.

**Synthesis of Au/RCC3.** The synthetic procedure used above to prepare Au@I-Cage-Cl was followed by using 9 mL of water solution containing RCC3 (15 mg) in place of I-Cage-Cl.

**Synthesis of Au/4-Cyanomethyl-1-vinyl-imidazolium bromide.** The

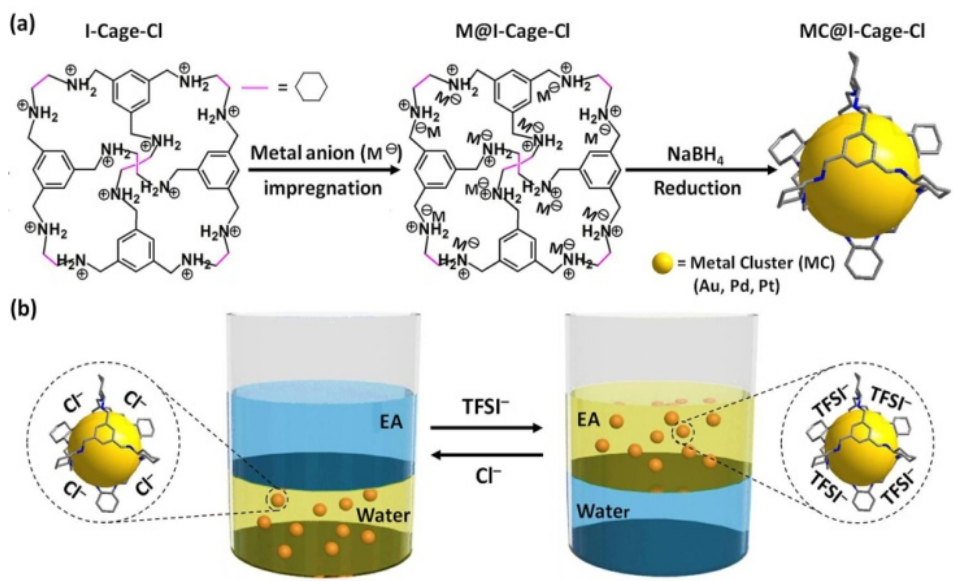

**Scheme 1.** (a) Synthetic scheme of MCs stabilized by I-Cage-Cl (Cl as counterion) in an aqueous phase (for clarity reasons Cl$^-$ anions are omitted; M$^-$ = [AuCl$_4$]$^-$, [PdCl$_4$]$^{2-}$ and [PtCl$_4$]$^{2-}$). (b) Anion exchange-driven phase transfer of I-Cage-X (X = Cl$^-$ or TFSI$^-$, TFSI = bis(trifluoromethane sulfonyl)imide)-capped MCs between aqueous and ethyl acetate (EA) phases.

**Synthesis of I-Cage-Cl.[12b]**

RCC3 (500 mg) was dissolved in $CHCl_3$ (10 ml) by stirring. Hydrogen chloride (in dioxane, 2.50 ml) was added dropwise. The precipitate appeared and the reaction mixture was stirred for a further 2 h at room temperature. The precipitate was collected by filtration then washed by $CHCl_3$ for several times to give solid powder after being dried under vacuum at 90 °C. $^1$H NMR (D$_2$O, 400 MHz) δ 7.62 (s, 12H, -ArH), 4.47 (d, 12H, -ArCH2), 4.23(d, 12H, -ArCH2), 3.69 (d, 12 H, CH on cyclohexane), 2.55-1.27 (m, 48H, CH$_2$ on cyclohexane) ppm (The details of peak integration are shown in Figure S35); $^{13}$C NMR (D$_2$O, 400 MHz): δ 133.0, 130.0, 58.5, 48.3, 25.8, 22.6 ppm. The reported RCC3 formula is $C_{72}H_{108}N_{12}$ (M = 1141.73),[12a] accordingly, the estimated molecular formula for cationic cage part is $C_{72}H_{120}N_{12}$ (the 12 -NH- units per cage are fully protonated by acid, M = 1154). This synthetic procedure used above to prepare Au@I-Cage-Cl was followed by using 9 mL of water solution containing 4-Cyanomethyl-1-vinyl-imidazolium bromide (15 mg) in place of I-Cage-Cl.

**Phase transfer of Au@I-Cage-X (X = Cl$^-$ or TFSI$^-$) driven by anion exchange.**

Typically, 1 mL of as-synthesized Au@I-Cage-Cl (above section) was layered with 1 mL ethyl acetate (EA), then 3 mg LiTFSI was added into the solution and rigorously shaking to let Au@I-Cage-Cl transfer to EA (Au@I-Cage-TFSI) (The addition of a few drop of methanol (∼50 μl) that is miscible in both water and EA can accelerate the transfer process). Then 6 mg KCl was added into the solution to achieve the reversible transfer to water solution.

**Synthesis of catalyst for AB hydrolysis reaction.**

**Synthesis of Pt@I-Cage-Cl catalyst.** The synthetic procedure used above to prepare Au@I-Cage-Cl was followed by using 6 mL of aqueous solution containing I-Cage-Cl (15 mg), 0.5 mL of aqueous solution containing $K_2PtCl_4$, (0.01 mmol Pt in content) in place of $HAuCl_4$.

**Synthesis of Pt/RCC3 catalyst.** The synthetic procedure used above to prepare Pt@I-Cage-Cl catalyst was followed by using 6 mL of aqueous solution containing RCC3 (15 mg) in place of I-Cage-Cl.

**Synthesis of Pt-Support-Free (Pt-SP-Free) catalyst.** The synthetic procedure used above to prepare Pt@I-Cage-Cl catalyst was followed by using 6 mL of aqueous solution without I-Cage-Cl stabilizer.

**Synthesis of Pt/CTAB catalyst.** The synthetic procedure used above to prepare Pt@I-Cage-Cl catalyst was followed by using 6 mL of aqueous solution containing CTAB (15 mg) in place of I-Cage-Cl.

**Synthesis of Pt/PVP catalyst.** The synthetic procedure used above to prepare Pt@I-Cage-Cl catalyst was followed by using 6 mL of aqueous solution containing PVP (30 mg) in place of I-Cage-Cl.

**Catalytic activity characterization**

**Procedure for the hydrolysis of AB by Pt@I-Cage-Cl catalyst:** The reaction apparatus for measuring the hydrogen evolution from the hydrolysis of AB is as follows. In general, the as-synthesized Pt@I-Cage-Cl catalyst was placed in a two-necked round-bottomed flask (30 mL), which was placed in a water bath under ambient atmosphere. A gas burette filled with water was connected to the reaction flask to measure the volume of hydrogen. The reaction started when AB (15 mg) in 0.5 ml water was added into the flask. The volume of the evolved hydrogen gas was monitored by recording the displacement of water in the gas burette. The reaction was completed when there was no more gas generation. The hydrolysis of AB can be expressed as follows:
$NH_3BH_3 + H_2O \rightarrow NH_3BO_2 + 3H_2$ (1)

**Procedures for the hydrolysis of AB by Pt/RCC3, Pt-SP-Free, Pt/CTAB and Pt/PVP catalysts:** The procedures for the hydrolysis of AB were similar to that of Pt/I-Cage-Cl catalyst except different catalysts were used.

## Results and discussion

The targeted ionic cage molecule (denoted as I-Cage-Cl, Cl as counterion), $[(H_{12}RCC3)^{12+}12Cl^-]$, was synthesized according to a recent report by HCl-acidifying a neutral amine cage RCC3 (Scheme S1).[12a,b] Each I-Cage-Cl carries 12 cations and was characterized by proton and carbon nuclear magnetic resonance ($^1H$ and $^{13}C$ NMR) spectra and mass spectrometry (Figure S1-S3). The hydrophilic ammonium chloride ion pair makes I-Cage-Cl readily water-soluble, while its neutral precursor RCC3 is insoluble. The cryogenic electron microscopy (cryo-EM) image (Figure S4) of I-Cage-Cl reveals well-dispersed dark dots of 3 ± 0.4 nm in size, slightly larger than its precursor RCC3 of ∼ 2 nm in size (cryo-EM image in Figure S5, which is in agreement with size estimation from its octahedron architecture.[12b]). The size expansion from RCC3 to I-Cage-Cl is expected as the ionized -NH- units electrostatically stretch the cage and the newly introduced $Cl^-$ also expand its overall hydrodynamic size.

Due to a high cation density on the cage, the Coulumbic attraction and the confinement effect facilitate encapsulation of metal anions into I-Cage-Cl as host, as reported recently for a host of metals on different supports.[13] Principally, the reduction of encapsulated metal anions and the subsequent nucleation within the cage host is expected and downsizes MCs to match the cage cavity. Experimentally this process runs as follows: a pale yellow solution of I-Cage-Cl (15 mg) in 9 mL of water was prepared. It became light brown upon addition of $HAuCl_4$ (Au content: 0.5 mg). The interaction between the cationic cage and $[AuCl_4]^-$ was monitored by Zeta potential analysis, in which an initial value of + 38.9 mV of I-Cage-Cl drops to +28.6 mV upon addition of $HAuCl_4$ due to specific adsorption/complexation of $[AuCl_4]^-$ by I-Cage-Cl. Addition of $NaBH_4$ at r.t. produced a pale brown, stable MC solution.

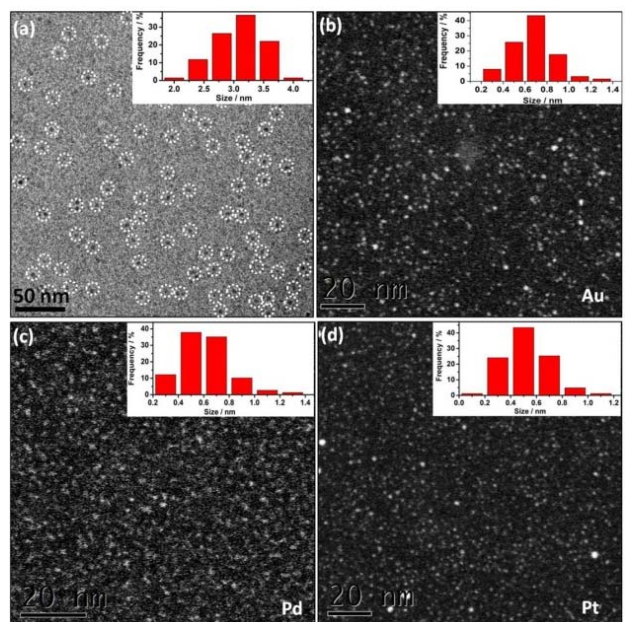

**Figure 1.** (a) Cryo-EM image of Au@I-Cage-Cl on a Lacey carbon grid and the size distribution histogram. Some Au@I-Cage-Cl are highlighted by white dotted circles. (b-d) HAADF-STEM images and the size distribution histograms of Au, Pd and Pt clusters, respectively.

Dynamic light scattering (DLS) study (Figure S6) and cryo-EM analysis (Figure 1a) determine the as-synthesized Au@I-Cage-Cl be to 3.0 nm and 3.1 ± 0.4 nm, respectively, in good consistence with the size value of native I-Cage-Cl. The size of Au clusters was determined by high-angle annular dark-field scanning transmission electron microscopy (HAADF-STEM) to be 0.65 ± 0.2 nm (Figures 1b and S7a), which matches well with the pore size of RCC3 cage (∼0.72 nm), indicative of cage encapsulation. Such small cluster size agrees well with the absence of a Plasmon peak in the UV-vis spectrum (Figure 2b). X-ray photoelectron spectroscopy (XPS) data identify peaks of binding energy at 88.4 and 84.4 eV, corresponding to $4f_{5/2}$ and $4f_{7/2}$ of Au clusters in the cage (Figure S8).

The NMR technique is applied to better analyze Au@I-Cage-Cl and specifically, the spatial relationship between the cage host and the Au cluster guest. The $^{13}C$ NMR spectra of Au@I-Cage-Cl (Figure S2b) and I-Cage-Cl (Figure S2a) show litter-to-no variation in peak positions, suggesting that I-Cage-Cl is in an approximate configuration before and after encapsulation of MCs. Interestingly, in comparison to I-Cage-Cl (Figure S1a), broadening of all peak widths in the $^1$H NMR spectrum of Au@I-Cage-Cl (Figure S1b) is observed, indicative of a local restricted motion and structural heterogeneity stemming from encapsulation of Au clusters.[14,12c] A similar observation was reported in Au NPs@Cage (average Au NP size:1.9 nm),[14] in which the cage shell tightly wrapped around the AuNP, and experienced restricted mobility and fast spin relaxation. Further evidence was given by 2D diffusion ordered spectroscopy $^1$H nuclear magnetic resonance (2D DOSY $^1$H NMR) (Figure S9&10) that shows similar diffusion coefficients for Au@I-Cage-Cl (2.16×10$^{-6}$ cm$^2$ s$^{-1}$) and I-Cage-Cl (2.06×10$^{-6}$ cm$^2$ s$^{-1}$), confirming the similar size and shape of native I-Cage-Cl and Au@I-Cage-Cl. Such observation proves that Au clusters rest inside the cage cavity instead of intercage interactions or aggregation on the cage surface.[14,12c] In addition, the size of Au@I-Cage-Cl determined by cryo-EM data in Figure 1a excludes the possibility of Au clusters stabilized by multiple cages.

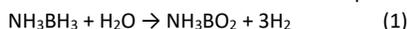

Cage-confinement brings Au clusters high stability. For example, storage in a temperature range of 77 to 363 K and in a pH range of 3 to 10 did not change the cluster size, as confirmed by UV-Vis spectra (Figure S11) and HAADF-STEM analysis (Figure S12-15). Rationally the electrostatic repulsion separates molecular cages and each octahedron cage sterically contains individual MCs well inside.[15] In a control experiment without I-Cage-Cl, only aggregated Au NPs were observed (Figure S16). Using neutral cage RCC3 as support produced Au NPs of a larger size of 4 ± 0.8 nm with a broader size distribution and poor solubility in water (Figure S17). Moreover, a monovalent ionic liquid (IL) 4-cyanomethyl-1-vinyl-imidazolium bromide, when used in the same experiment, resulted in Au NPs of 6 ± 0.9 nm in size (Figure S18). Therefore, the cage-ion synergy plays a pivotal role in size control. It is worth noting that our method can produce other noble MCs, i.e. 0.6 ± 0.2 nm for Pd (Figure 1c&S7b) and 0.5 ± 0.2 nm for Pt clusters (Figure 1d&S7c).

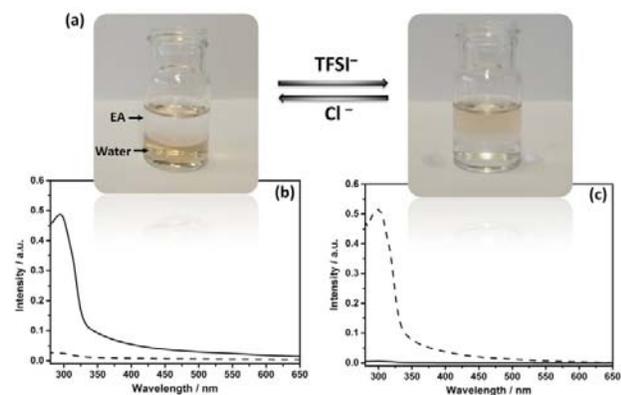

**Figure 2**. (a) Photographs showing reversible transfer of Au-I-Cage-X (X = Cl$^-$ or TFSI$^-$) between aqueous and EA phases upon anion exchange. (b, c) UV-Vis spectra of Au clusters transferred between aqueous and EA phases (solid curve for aqueous phase, dotted curve for EA phase) as shown in (a).

In IL chemistry, hydrophilicity/hydrophobicity of ILs is switchable by ion metathesis. Here, as a proof of concept, Au@I-Cage-Cl was modified into hydrophobic Au@I-Cage-TFSI bearing bis(trifluoromethane sulfonyl)imide (TFSI) anion by anion exchange with LiTFSI (Scheme 1b). Experimentally, hydrophobic ethyl acetate (EA) was first added to a pale brown Au@I-Cage-Cl aqueous solution to form a clear supernatant. Addition of LiTFSI formed a colorless aqueous phase and a pale brown EA phase (Figure 2a). Residue Au@I-Cage-Cl in the aqueous phase produced a negligible absorbance peak at λ = 300 nm (assigned to I-Cage) in its UV-Vis spectrum, while the EA phase shows a characteristic absorbance spectrum of I-Cage, indicative of a quantitative phase transfer (Figure 2c). A similar behavior was observed when Au clusters were reversed to aqueous phase by adding KCl (Figure 2b). Inductively coupled plasma atomic emission spectroscopy (ICP-OES) measurement shows < 5 ppm loss of Au (Figure S19) during this cycle, indicative of its reversible nature. Such process can be repeated three times without significantly altering the intensity and position of cage peak in UV-vis spectrum (Figure S20). Additionally, DLS (Figure S21) and HAADF-STEM analyses did not show any size increase (Figure S22a-c) for Au cluster, as also for Pd and Pt (Figure S22d-i,23&24). Worth to mention is that such feature is valuable, as it combines the physical confinement effect with ion exchange capability, which fails in conventional organic ligands.

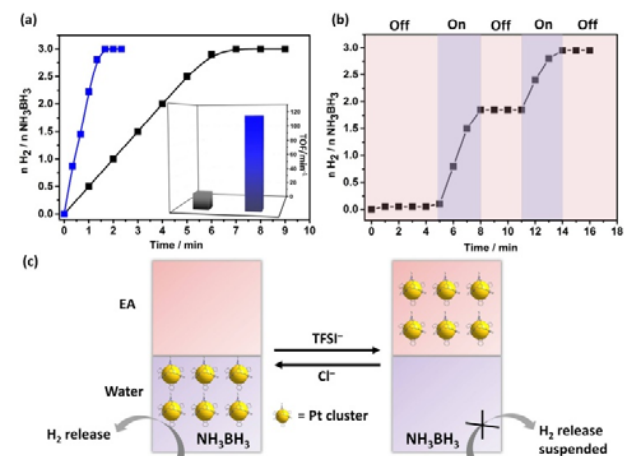

**Figure 3**. (a) Time course plots of H$_2$ generation for the hydrolysis of ammonia borane (AB) over the Pt@I-Cage-Cl (blue line) and unsupported Pt catalysts, Pt-SP-Free (dark line) at 300 K (Pt/AB = 0.02). Inset: the corresponding TOF values. (b) Control of H$_2$ release in the AB hydrolysis in water-EA mixture (pink and violet regions for "off" and "on" state, respectively). (c) Diagram of controlled H$_2$ release process.

Being physically confined and phase transferable is useful, as it allows the same MC catalyst for operations in different liquids while strictly keeping their size. Tested in a H$_2$ release reaction by hydrolysis of AB (NH$_3$BH$_3$) compound,[16] Pt@I-Cage-Cl at a Pt loading content of 0.01 mmol (Pt/AB = 0.02) completed the reaction (H$_2$/AB = 3.0) within 1.3 min at 300 K (Figure 3 and S25). The turnover frequency (TOF) is 115 min$^{-1}$, which is much higher than conventional surfactant-protected catalysts (Figure S26-29), e.g. Pt/CTAB (2.7 ± 0.3 nm, TOF: 5 min$^{-1}$), Pt/PVP (4 ± 0.4 nm, TOF: 13 min$^{-1}$), and RCC3-supported Pt NPs (Pt/RCC3, 3 ± 0.6 nm, TOF: 30 min$^{-1}$) (Figure S30&31), and comparable to the reported highly active monometallic Pt cluster/nanoparticle catalysts.[17] Not surprising, a low TOF value of 20 min$^{-1}$ was found for cage-free Pt nanoparticles (Pt-SP-Free) (Figure 3 and S32). Although I-Cage-Cl itself is catalytically inactive to AB (tested in Figure S33), the discrete open structure can promote catalytic reactions by downsizing Pt clusters and maximizing metal-reactant contact in solution.

The phase-transfer process is useful to recover and recycle catalyst. Typically, after completion of a reaction, Pt clusters were transferred to the EA phase by LiTFSI. The impure aqueous phase was replaced by clean water, before Pt clusters moved back by adding KCl for the next run (Figure S34). Uniquely, the anion exchange could regulate the $H_2$ production. As shown in Figure 3b&3c, $H_2$ generation was stopped upon addition of LiTFSI to evacuate Pt clusters from water; the reaction re-started by adding KCl to transfer Pt clusters back. This "on"/"off" cycle can be repeated till AB is consumed. Our system provides a new insight toward kinetic control for on-demand $H_2$ release in practical use.

## Conclusions

In conclusion, a method to prepare sub-nano-sized noble MCs hosted by ionic organic cage was reported. The reversible anion exchange process enables confined MCs to be shuttled selectively between an aqueous and an organic phase, which allows for efficient separation and recycling of MC catalysts from products or reaction mixtures. This method will equip researchers with a new tool to engineer MCs without compromising their catalytic activity, thus improving and augmenting applications in different research areas.

## Conflicts of interest

There are no conflicts to declare.

## Acknowledgements

J.Y. acknowledges financial support from the ERC Starting Grant (639720-NAPOLI) and the Wallenberg Academy Fellowship program KAW2017.0166 from the Knut and Alice Wallenberg Foundation. The authors acknowledge the National Natural Science Foundation of China (Grant Nos.21573016/21403241).

## Notes and references

# Supporting Information

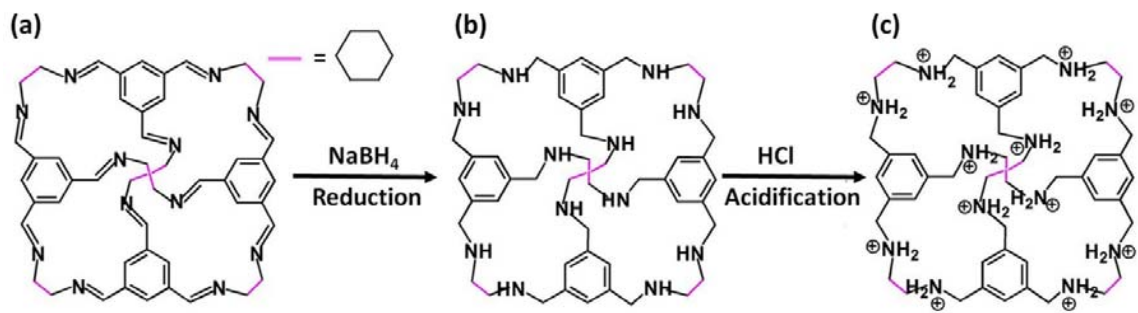

**Scheme S1.** Scheme illustrating the procedure to prepare the ionic organic cage (I-Cage) from its CC3 precursor, (a) CC3, (b) R-CC3, and (c) I-Cage-Cl (The Cl⁻ counter anions are omitted for clarity).

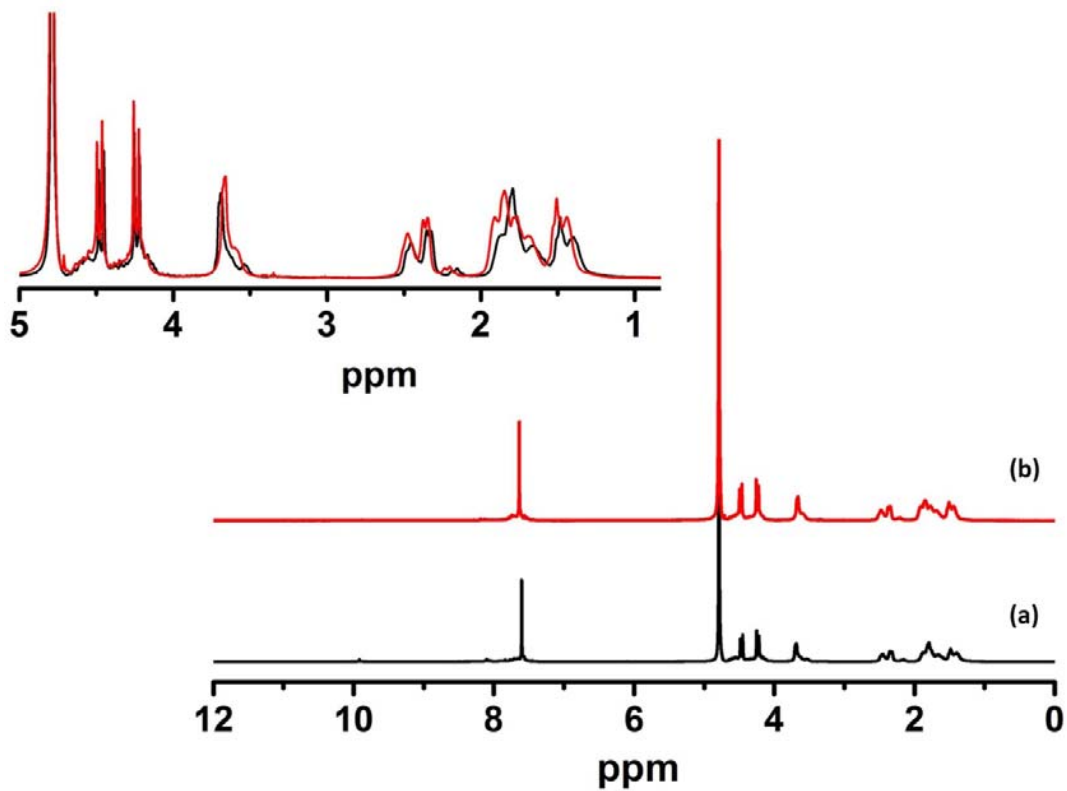

**Figure S1**. Chemical structures and $^1$H-NMR spectra of (a) I-Cage-Cl (The integration of the spectrum can be referred in Figure S38) and (b) Au@I-Cage-Cl in D$_2$O. The inset on the top left shows the comparison between I-Cage-Cl and Au@I-Cage-Cl.

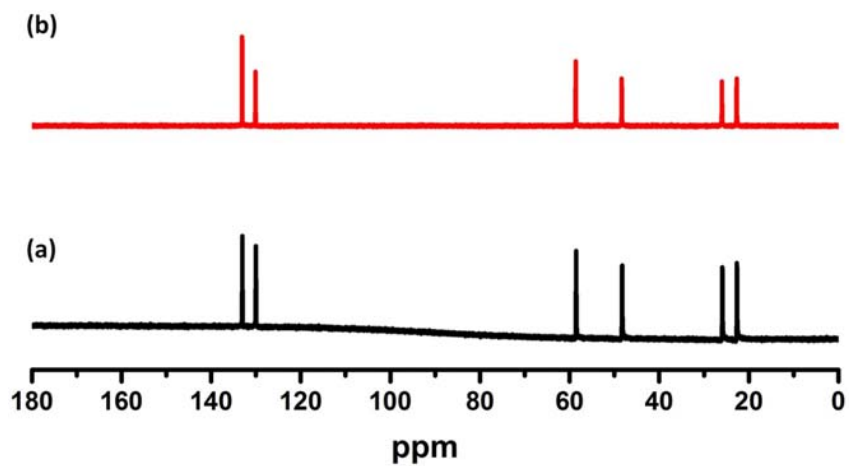

**Figure S2.** $^{13}$C-NMR spectra of (a) I-Cage-Cl and (b) Au@I-Cage-Cl in D$_2$O.

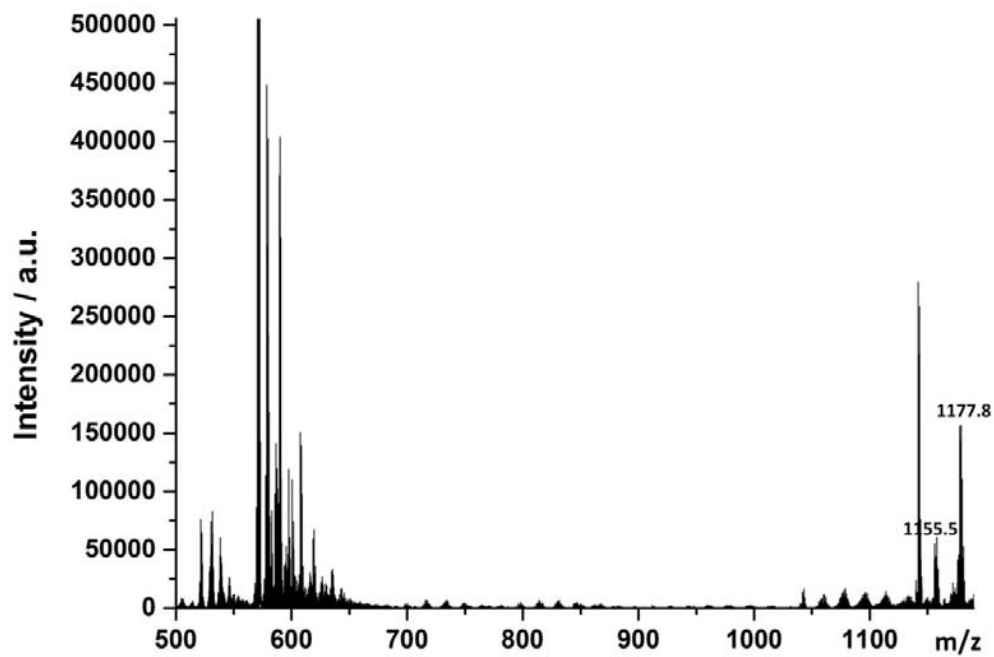

**Figure S3.** ESI(+)-MS spectrum of I-Cage-Cl, selected [M+H]$^+$ at m/z 1155.5 and selected [M+Na]$^+$ at m/z 1177.8 are marked in the picture.

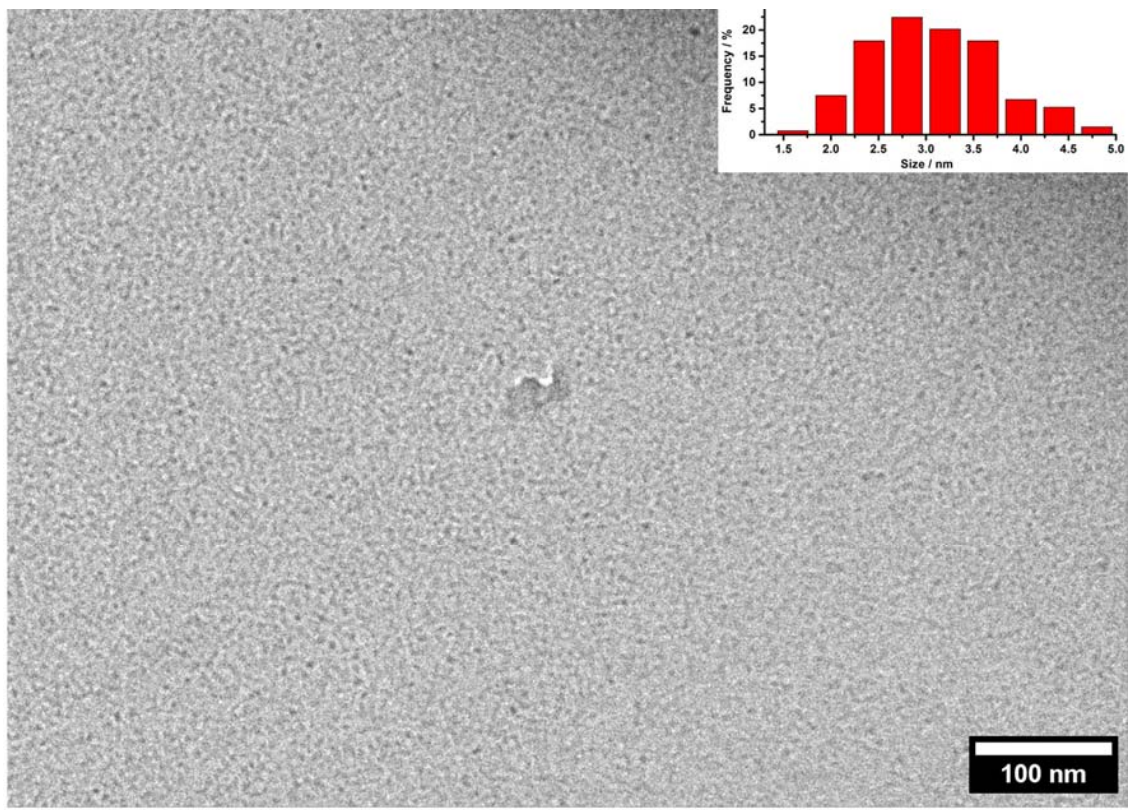

**Figure S4**. Cryo-EM image of the I-Cage-Cl (dark dot) on a Lacey carbon grid. The inset is the corresponding size distribution histogram of I-Cage-Cl.

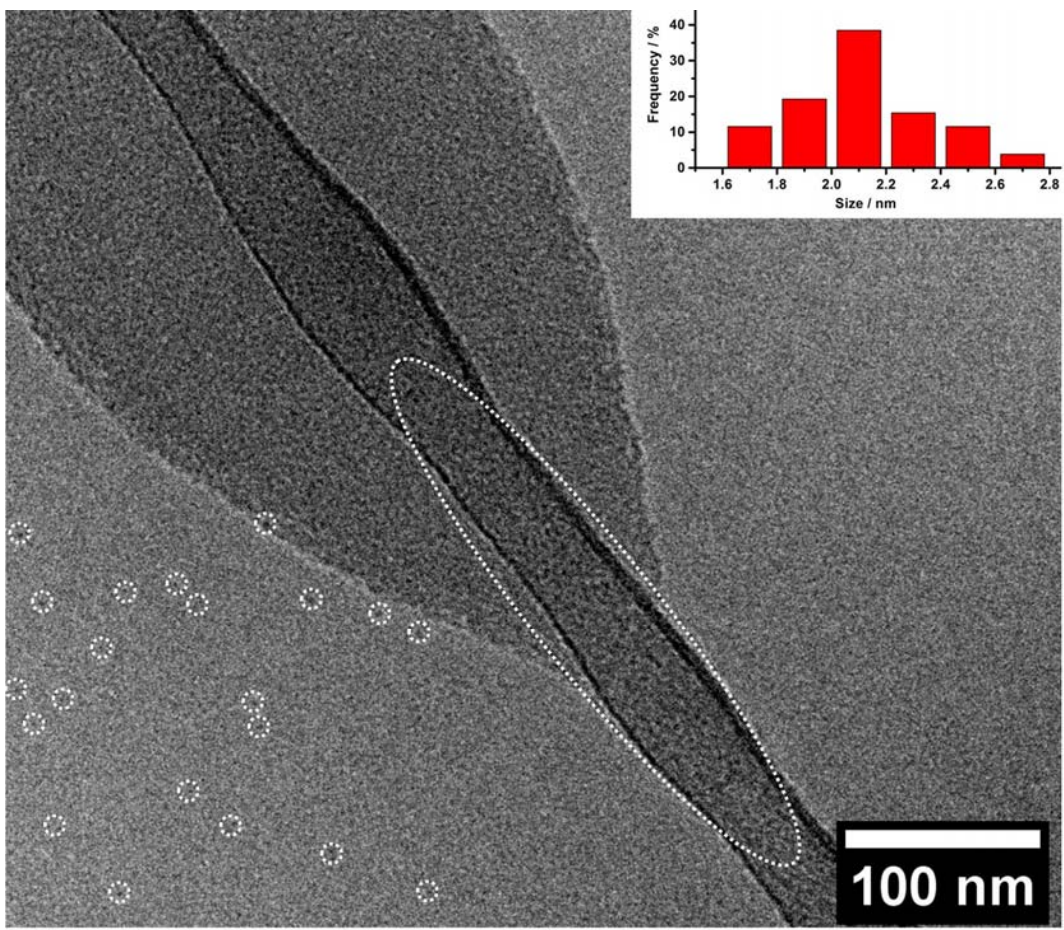

**Figure S5.** Cryo-EM image of the neutral RCC3 Cage (some dots highlighted by white circles) on a Lacey carbon grid. There are also a large number of RCC3 cage on Lacey carbon (highlighted by white dotted ellipses). The inset is the corresponding size distribution histogram of RCC3 cage.

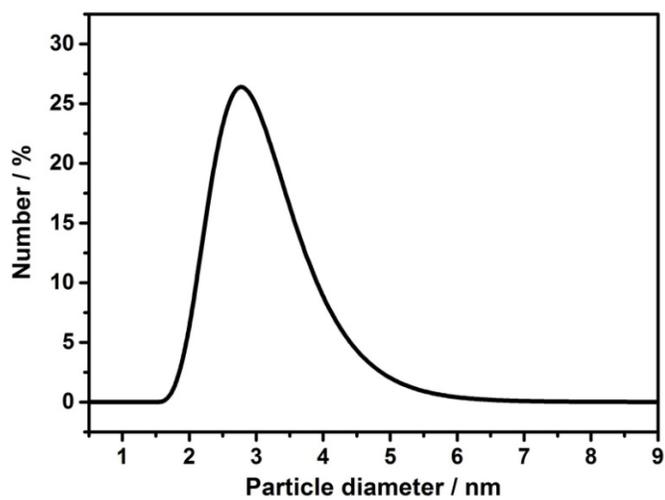

**Figure S6.** The number-average size distribution of the Au@I-Cage-Cl in aqueous solution observed by DLS.

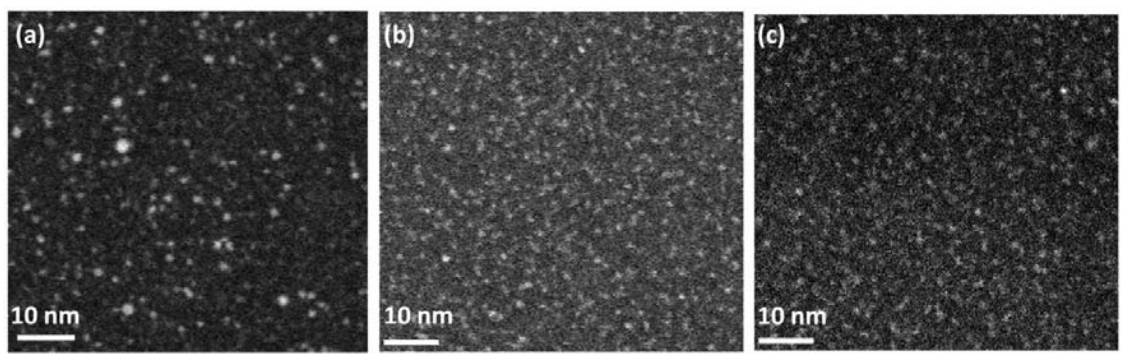

**Figure S7.** The HAADF-STEM images of (a) Au, (b) Pd and (c) Pt clusters in aqueous solution.

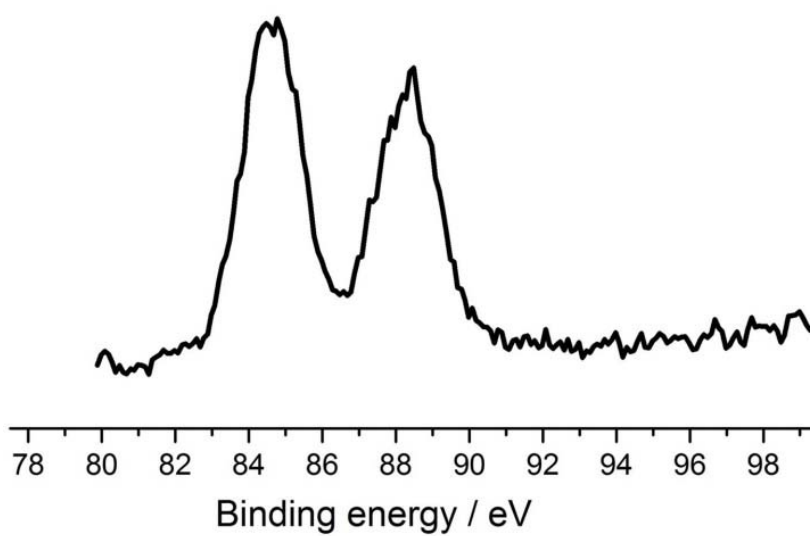

**Figure S8.** XPS spectrum of Au@I-Cage-Cl showing Au 4f$_{7/2}$ (84.4 eV) and 4f$_{5/2}$ (88.4 eV) peaks of metallic Au.

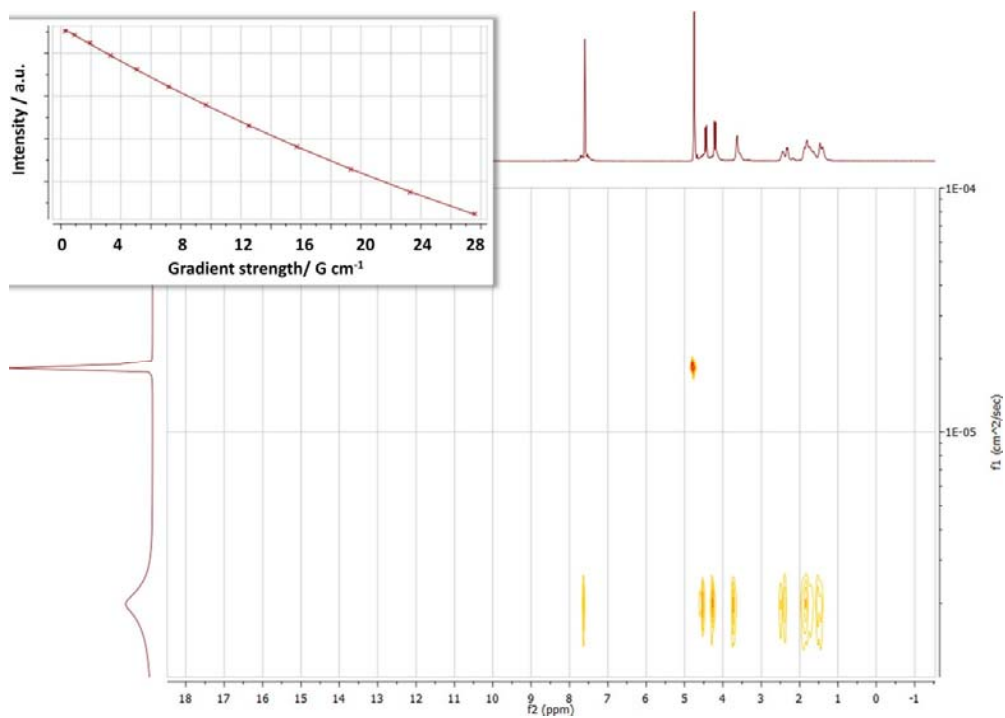

**Figure S9.** $^1$H 2D-DOSY NMR spectrum of I-Cage-Cl in D$_2$O. The inset in upper left is the plot of the signal intensity as a function of the gradient strength. Diffusion coefficients are obtained by non-linear fitting of the decay curve.

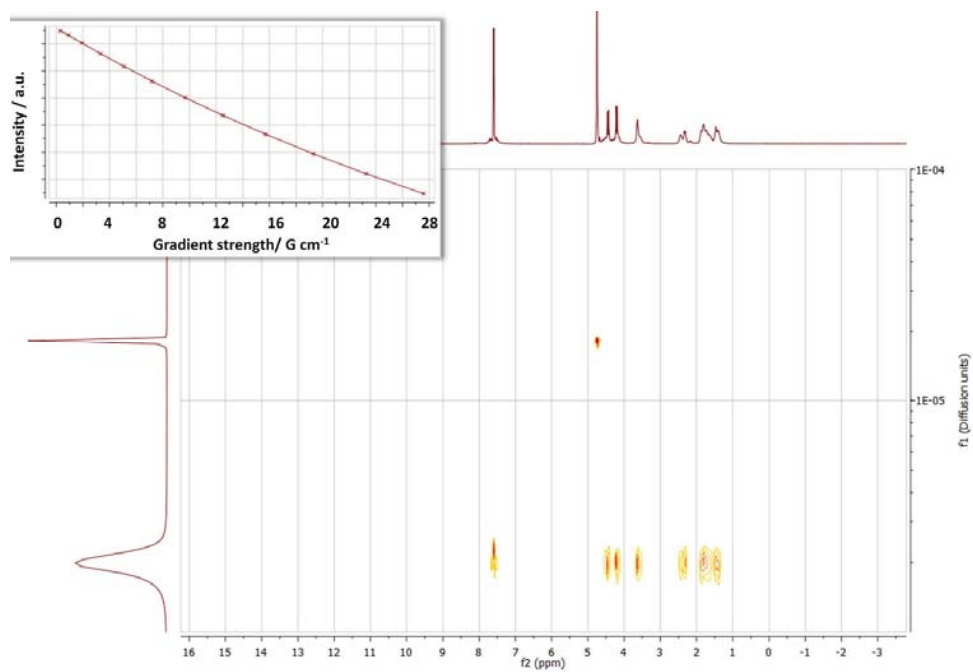

**Figure S10.** $^1$H 2D-DOSY NMR spectrum of Au@I-Cage-Cl in D$_2$O. The inset in upper left is the plot of the signal intensity as a function of the gradient strength. Diffusion coefficients are obtained by non-linear fitting of the decay curve.

**Note**: The two diffusion coefficients were calculated using the peaks at 7.62 ppm for I-Cage-Cl and Au@I-Cage-Cl. These two peaks are well separated from other resonances, so they are described by a mono-exponential function. Data were analysed by plotting the signal intensities (areas) as a function of the gradient strength, followed by non-linear fitting of the resulting decay curves. For internal consistency, we checked the method by calculating the diffusion for $D_2O$, which was used as a solvent, from the peak at 4.79 ppm: the diffusion coefficient value $D(D_2O) = 1.84 \times 10^{-5}$ cm$^2$ s$^{-1}$ obtained is in agreement with that of $D_2O$ reported in literature.(*J. Phys. Chem.*, 1965, 69, 4412–4412). The diffusion coefficient for I-Cage-Cl was measured as $D(\text{I-Cage-Cl}) = 2.06 \times 10^{-6}$ cm$^2$ s$^{-1}$ and for Au@I-Cage-Cl as $D(\text{Au@I-Cage-Cl}) = 2.16 \times 10^{-6}$ cm$^2$ s$^{-1}$.

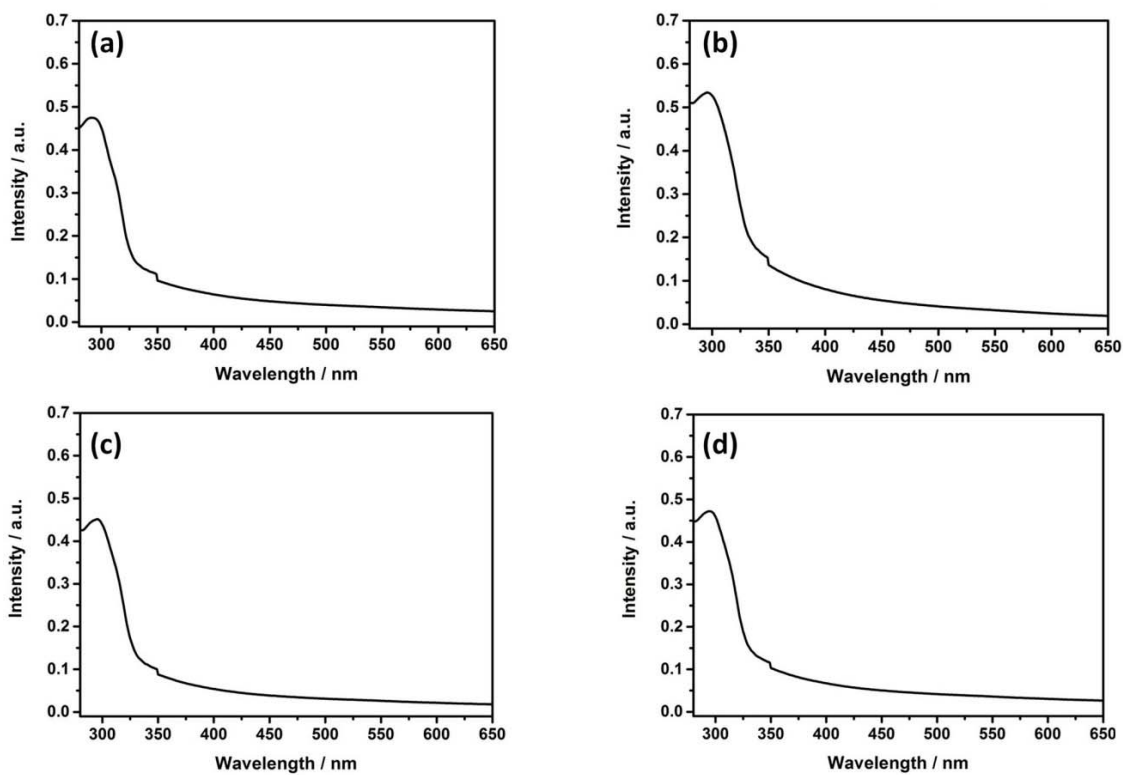

**Figure S11**. UV-vis spectra of Au cluster after staying in (a) acid, (b) base, (c) heat-treatment at 363 K and (d) liquid $N_2$.

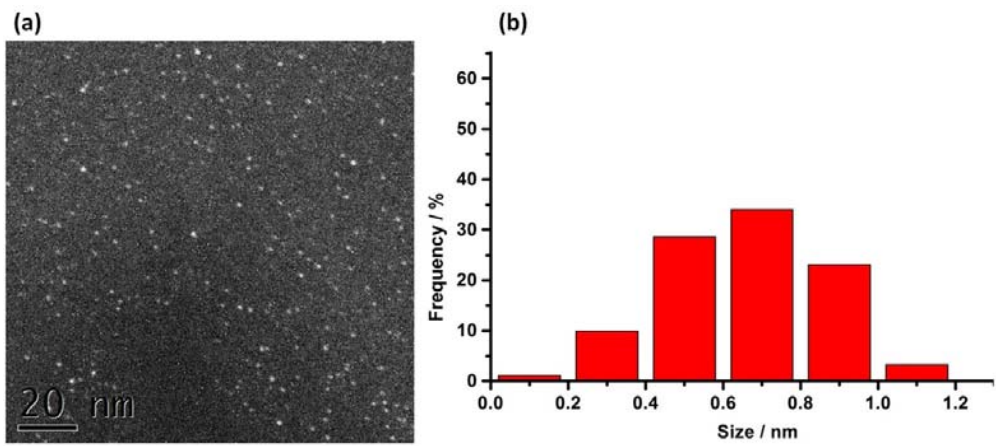

**Figure S12**. (a) HAADF-STEM image of Au clusters in an acid solution (pH=3) and (b) the corresponding size distribution histogram of Au clusters.

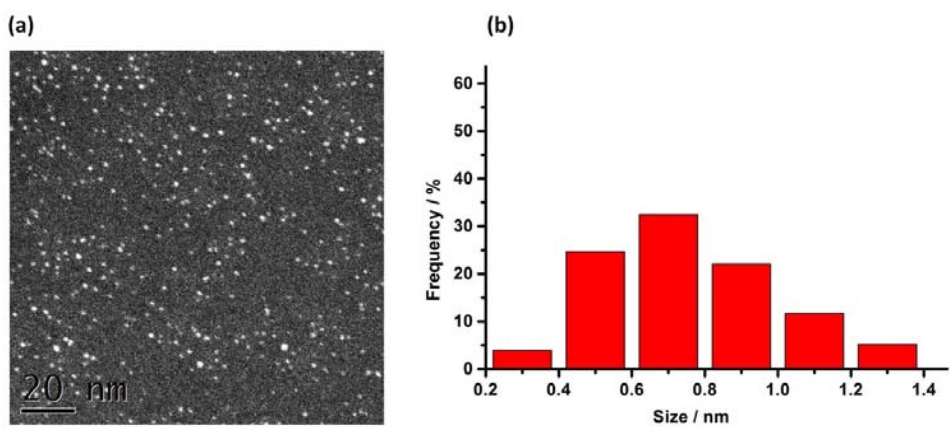

**Figure S13**. (a) HAADF-STEM image of Au clusters in a base solution (pH=10) and (b) the corresponding size distribution histogram of Au clusters.

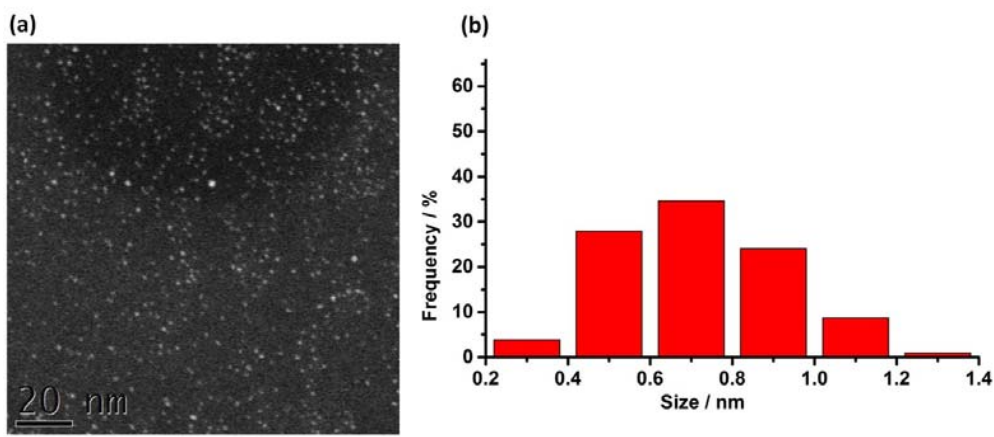

**Figure S14**. (a) HAADF-STEM image of Au clusters after treated by liquid nitrogen and (b) the corresponding size distribution histogram of Au clusters.

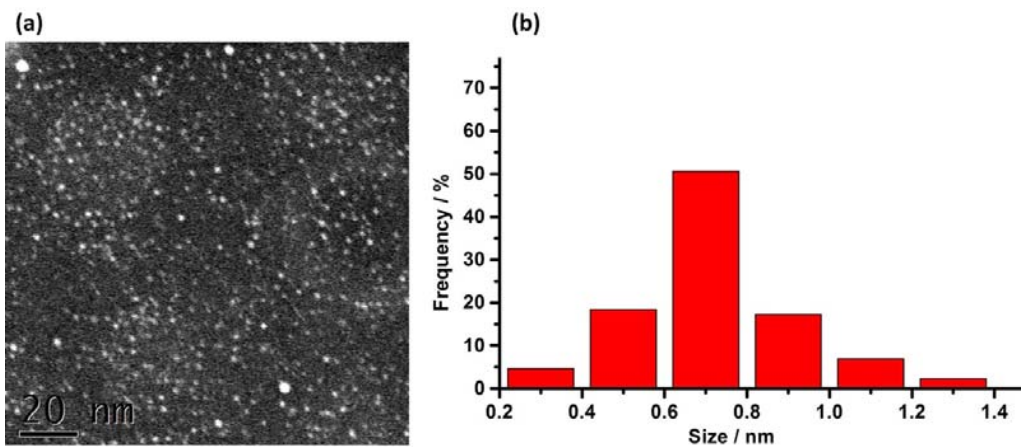

**Figure S15**. (a) HAADF-STEM images of Au cluster after heat-treatment at 363 K and (b) the corresponding size distribution histogram of Au clusters.

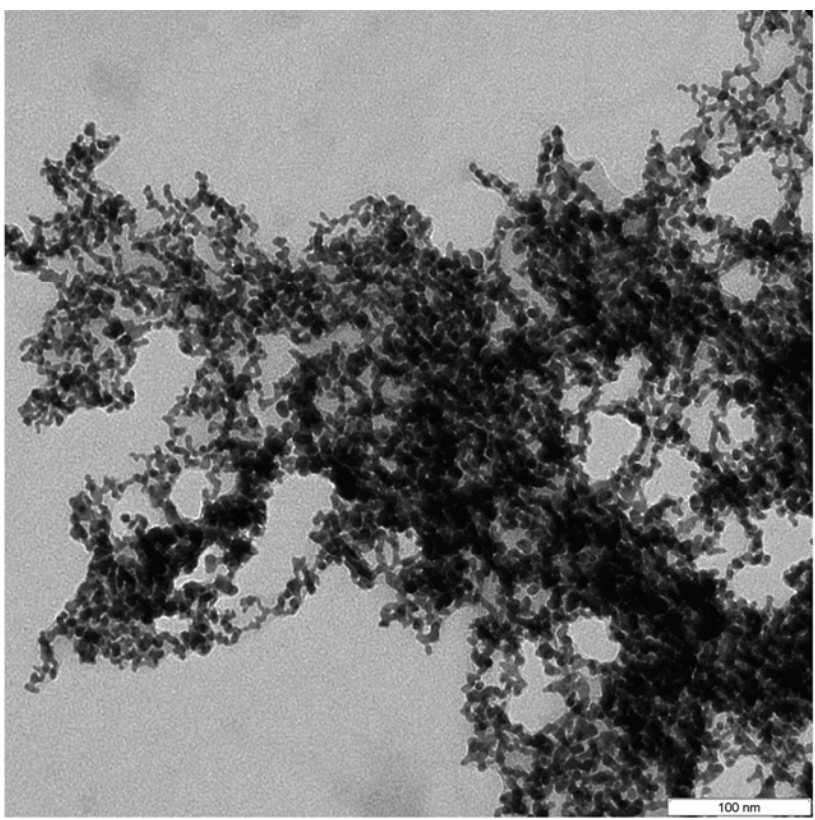

**Figure S16.** TEM image of Au nanoparticles produced without any support.

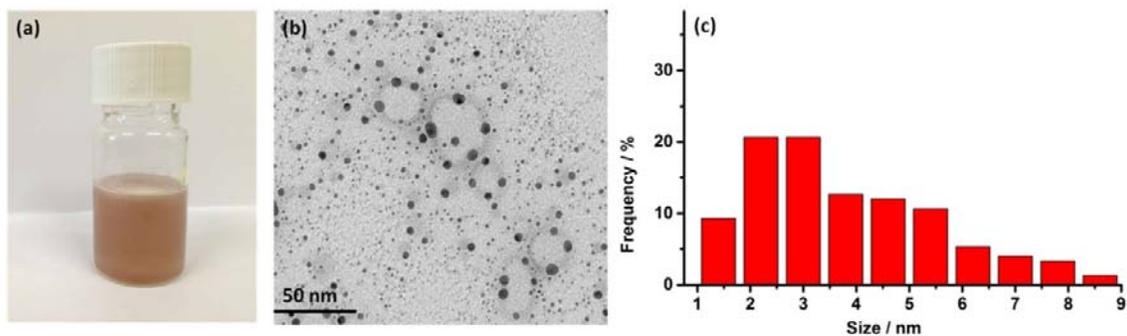

**Figure S17**. (a) Photograph of Au/RCC3 solution. (b) TEM image of Au/RCC3 and (c) the corresponding size distribution histogram of Au nanoparticles (4 ± 0.8 nm).

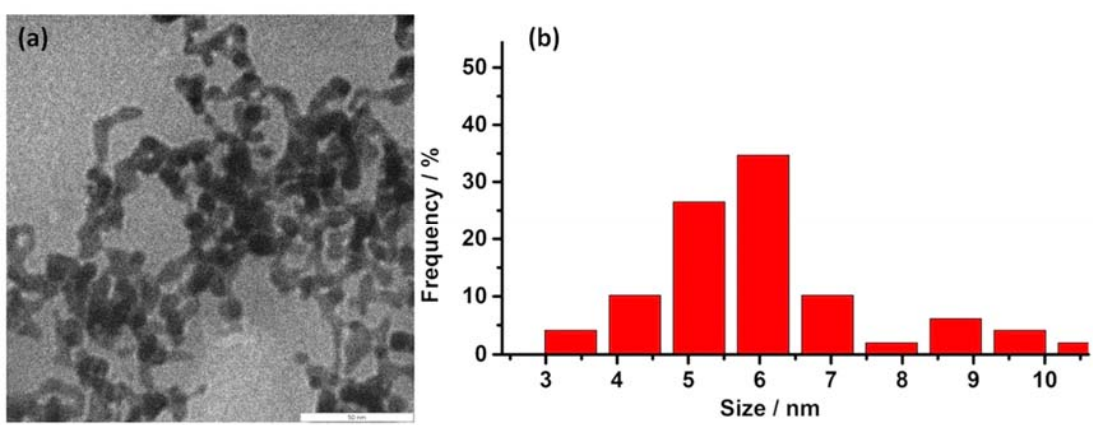

**Figure S18.** (a) TEM image of Au/4-cyanomethyl-1-vinyl-imidazolium bromide and (b) the corresponding size distribution histogram of Au nanoparticles (6 ± 0.9 nm).

| Phase transfer process | Amount of Au (mg/L, ppm) left in mother phase |
|---|---|
| From aqueous to EA phase | 4.5 (in aqueous phase) |
| From EA to aqueous phase | <1 (in EA phase) |

**Figure S19**. ICP-OES result of Au metal left in the mother phase after phase-transfer process.

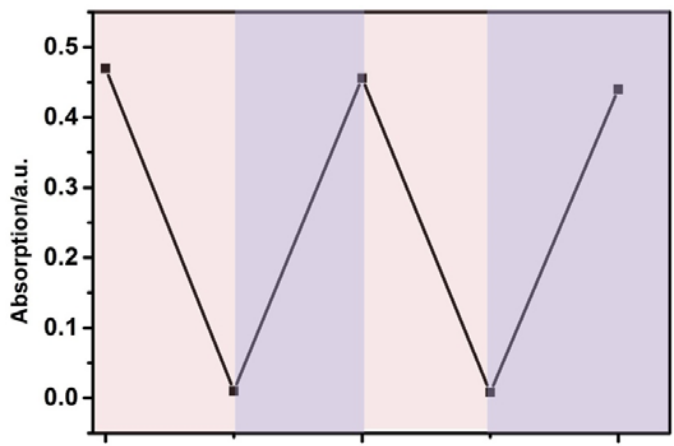

**Figure S20.** Reversible transfer of Au-I-Cage-Cl (absorbance at 300 nm) in water upon alternating addition of LiTFSI (pink rectangle) and KCl (violet rectangle).

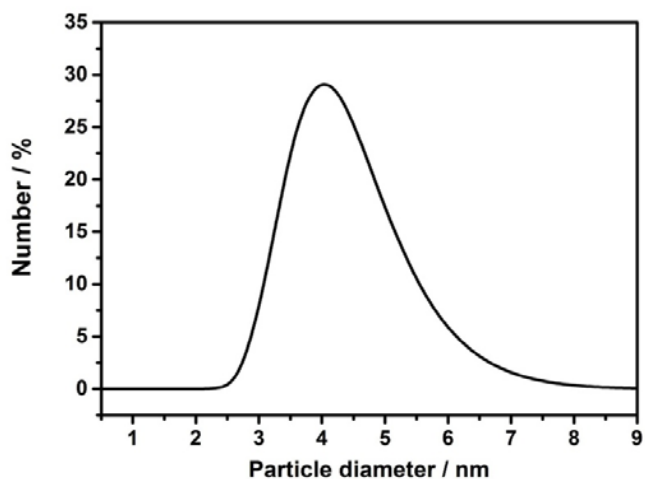

**Figure S21**. The number-average size distribution of the Au@I-Cage-TFSI in EA solution observed by DLS.

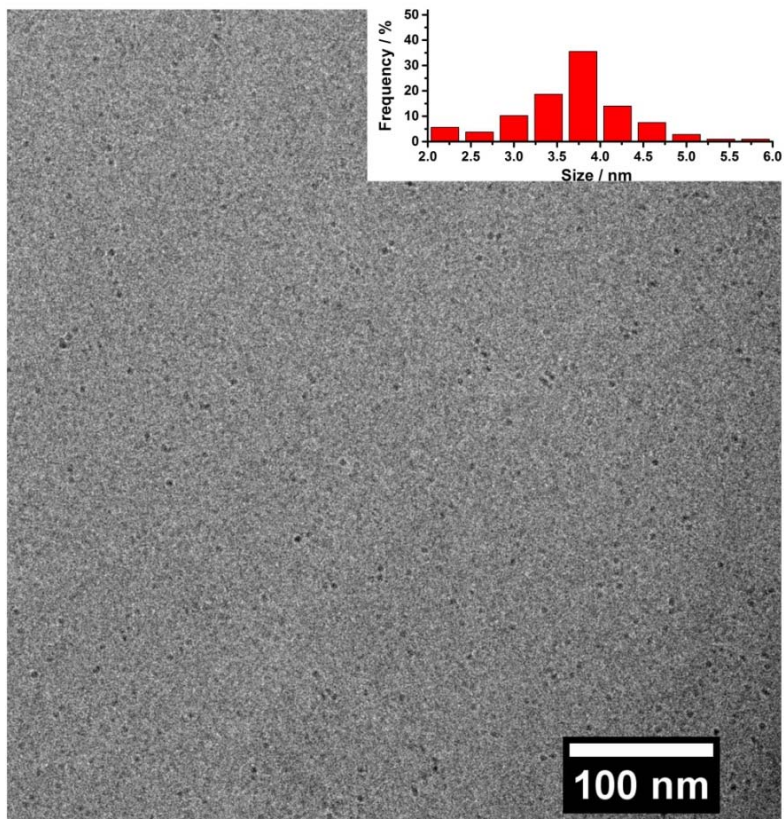

**Figure S22**. Cryo-EM image of Au@I-Cage-TFSI on a Lacey carbon grid and the size distribution histogram as inset.

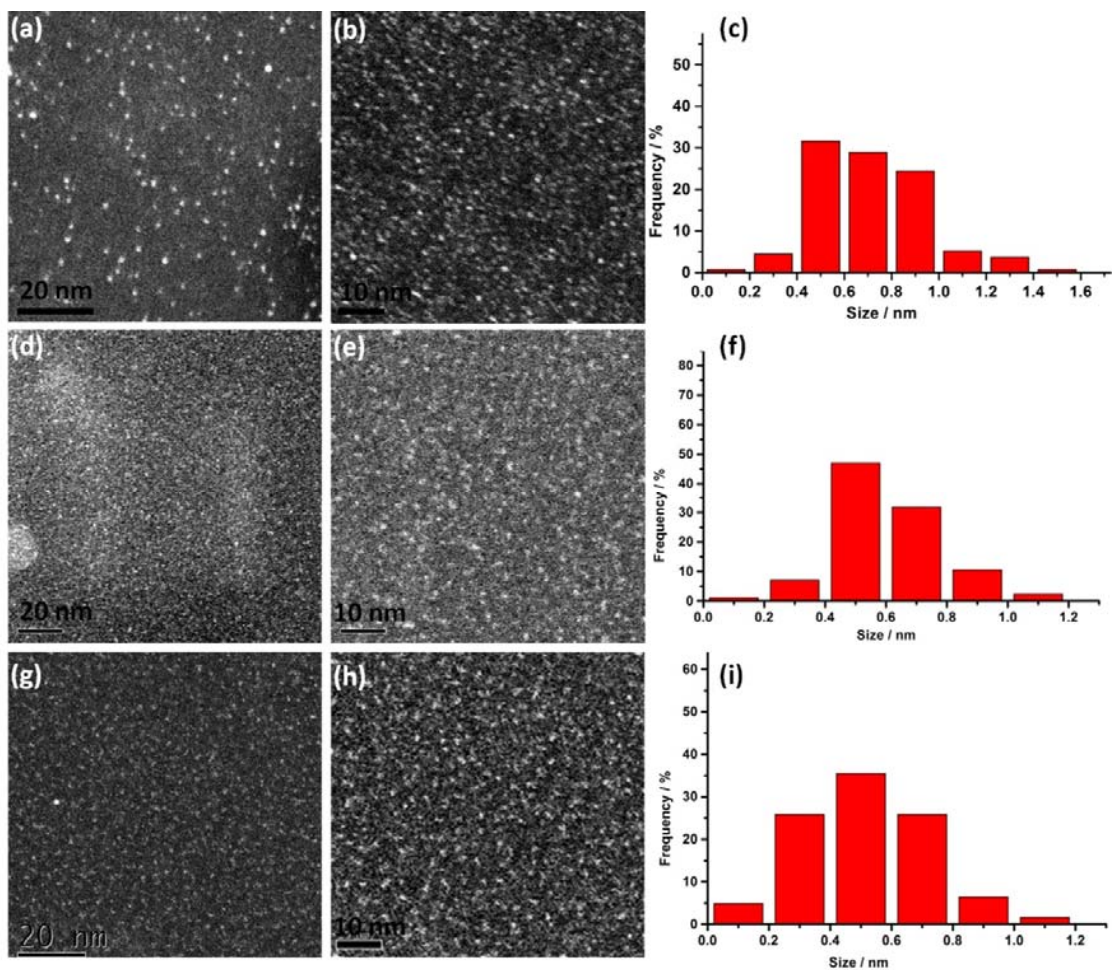

**Figure S23**. HAADF-STEM images at different magnifications and their corresponding size distribution of Au (a-c), Pd (d-f) and Pt (g-i) clusters in EA solution.

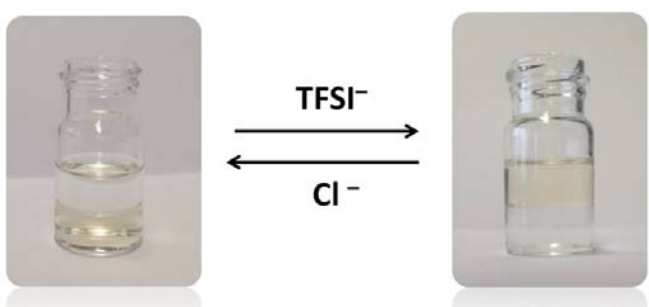

**Figure S24.** Photographs of the reversible phase transfer of Pd clusters assisted by the cage molecules between an aqueous and EA phases upon anion exchange.

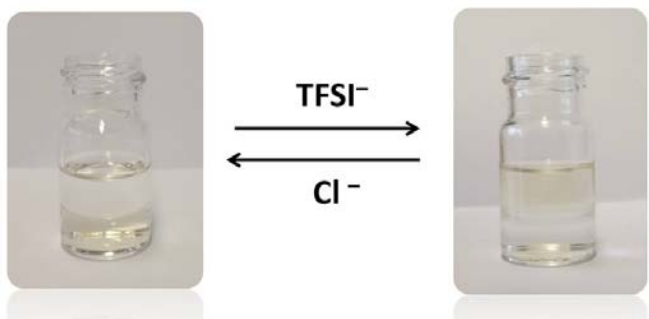

**Figure S25**. Photographs of the reversible phase transfer of Pt clusters assisted by the cage molecules between an aqueous and EA phases upon anion exchange.

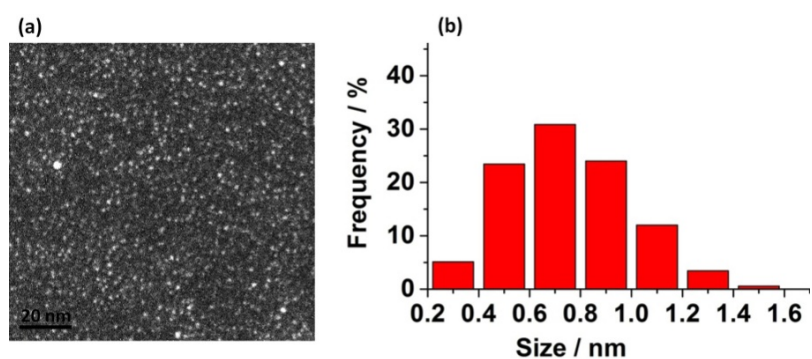

**Figure S26**. (a) HAADF-STEM image of Pt@I-Cage-Cl catalyst and (b) the corresponding size distribution histogram of Pt clusters (0.75 ± 0.2 nm).

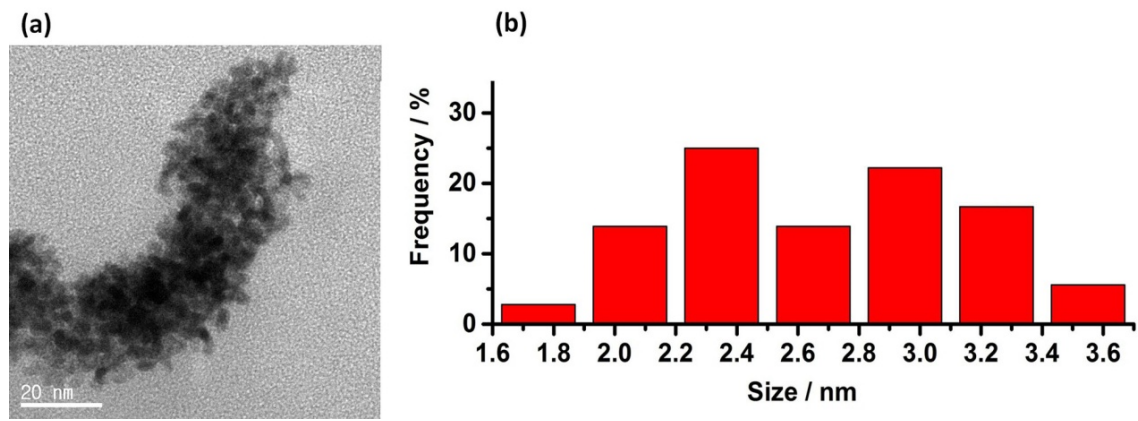

**Figure S27**. (a) TEM image of Pt/CTAB and (b) the corresponding size distribution histogram of Pt nanoparticles (2.7 ± 0.3 nm).

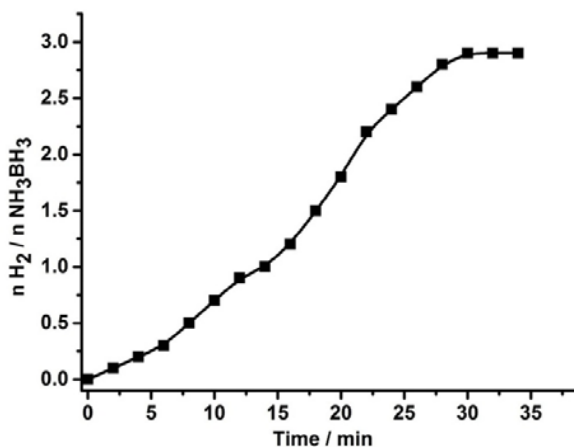

**Figure S28**. The time course plot of $H_2$ generation for Pt/CTAB catalyst.

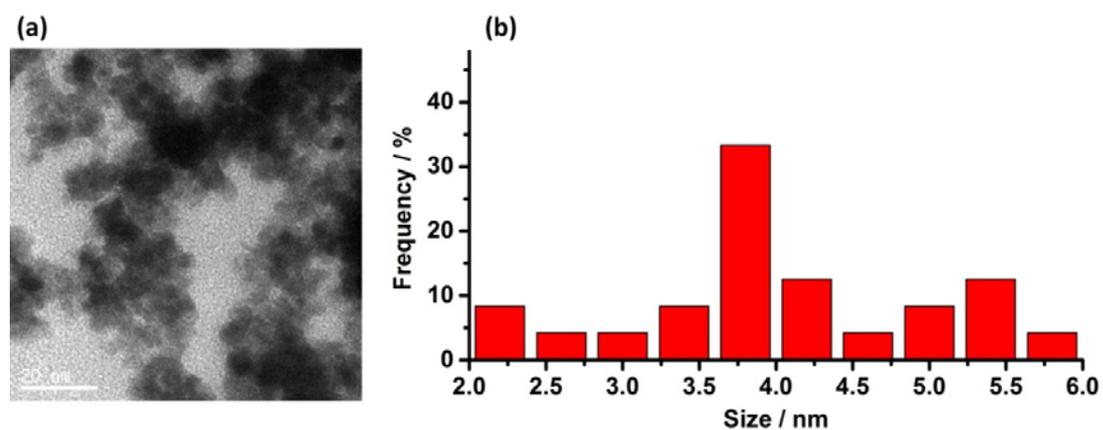

**Figure S29**. (a) TEM image of Pt/PVP and (b) the corresponding size distribution histogram of Pt nanoparticles (4 ± 0.4 nm).

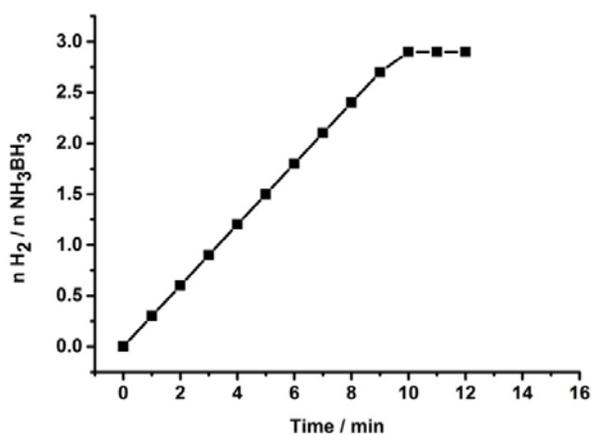

**Figure S30**. The time course plot of $H_2$ generation for Pt/PVP catalyst.

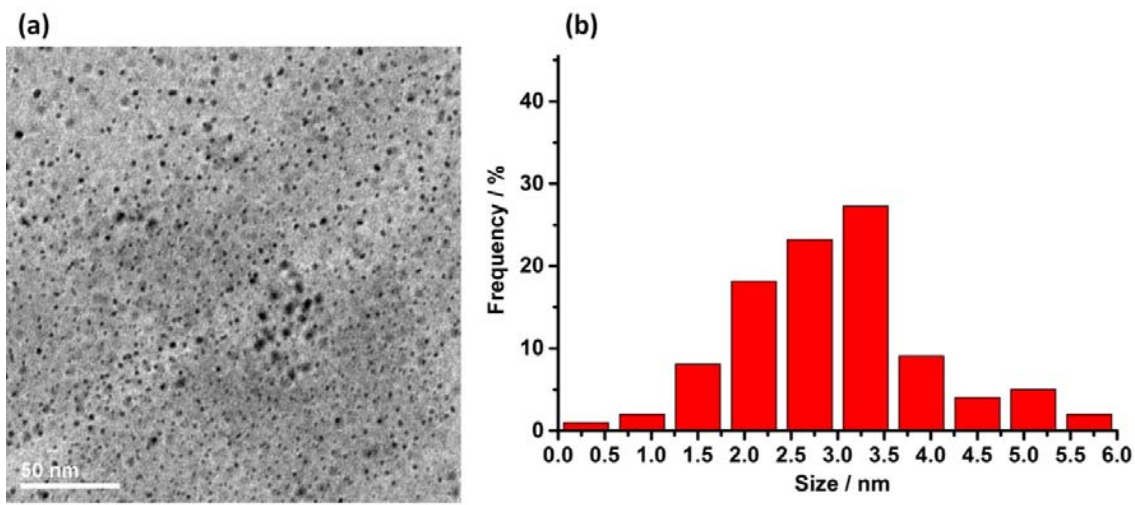

**Figure S31.** (a) TEM image of Pt/RCC3 and (b) the corresponding size distribution histogram of Pt nanoparticles (3 ± 0.6 nm).

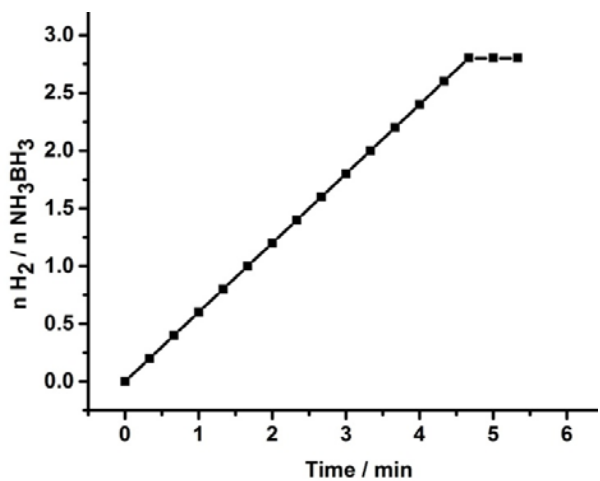

**Figure S32.** The time course plot of $H_2$ generation for Pt/RCC3 catalyst.

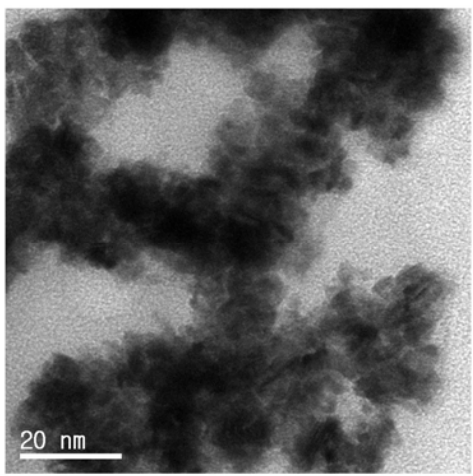

**Figure S33.** The TEM image of Pt-SP-Free catalyst.

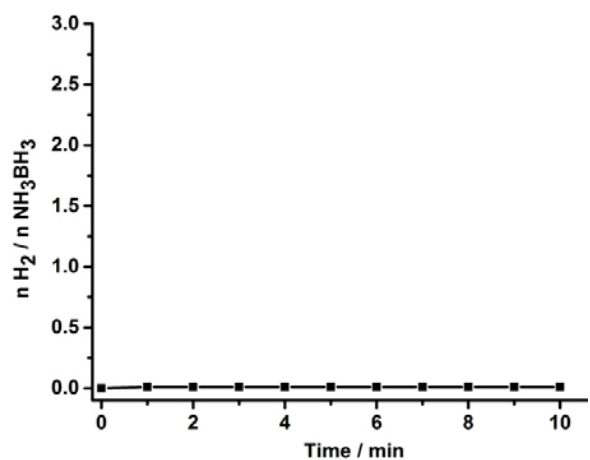

**Figure S34**. The time course plot of H$_2$ generation for pure I-Cage-Cl.

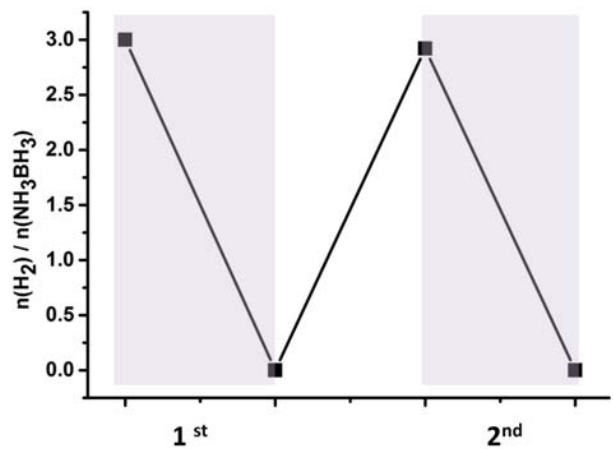

**Figure S35**. Recyclable AB hydrolysis reaction by Pt@I-Cage-Cl through anion exchange driven phase transfer.

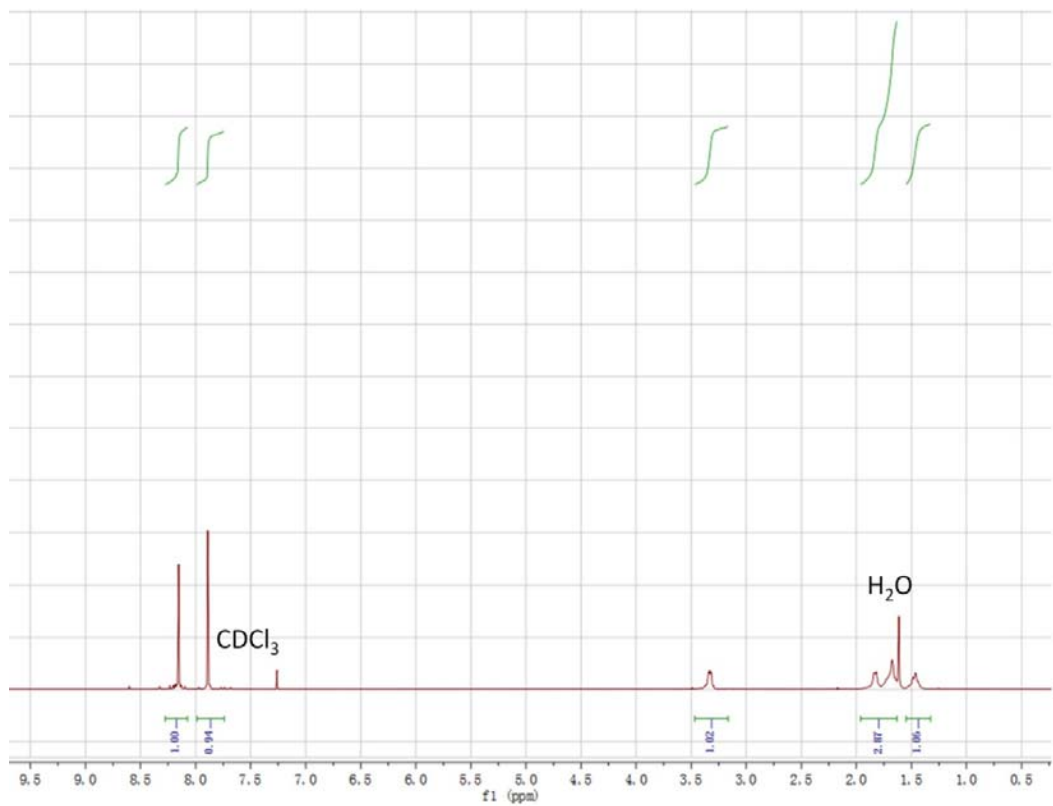

**Figure S36**. The integration of the $^1$H NMR spectrum of the CC3 in CDCl$_3$.

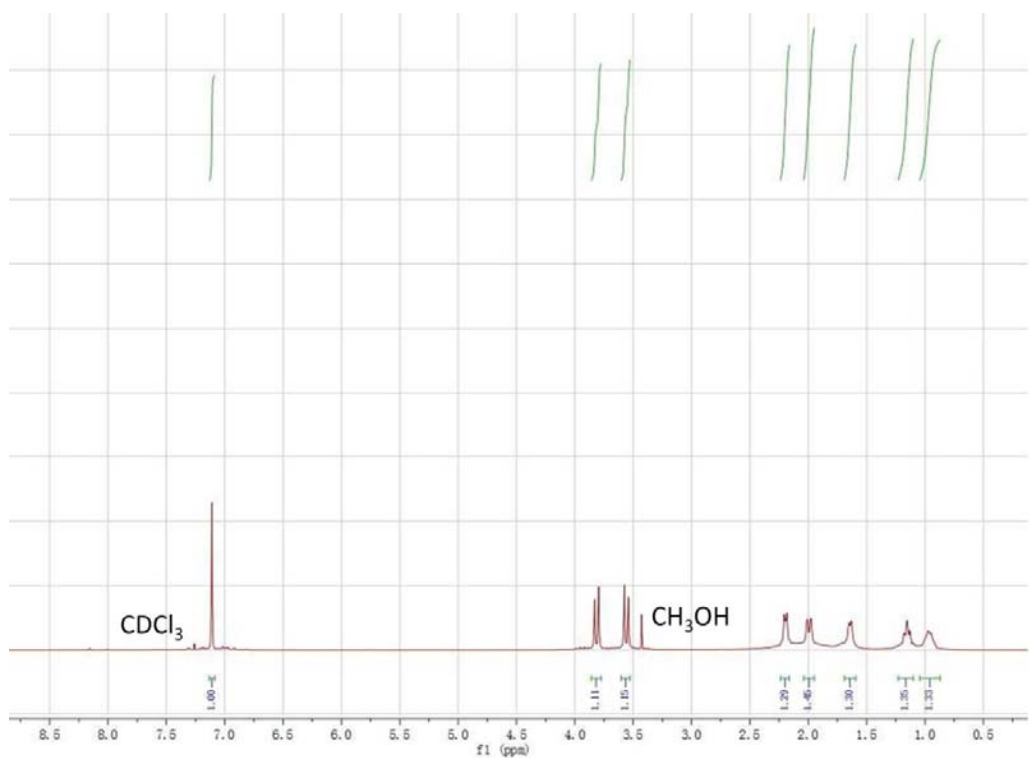

**Figure S37**. The $^1$H NMR spectrum of the RCC3 in CDCl$_3$.

S18

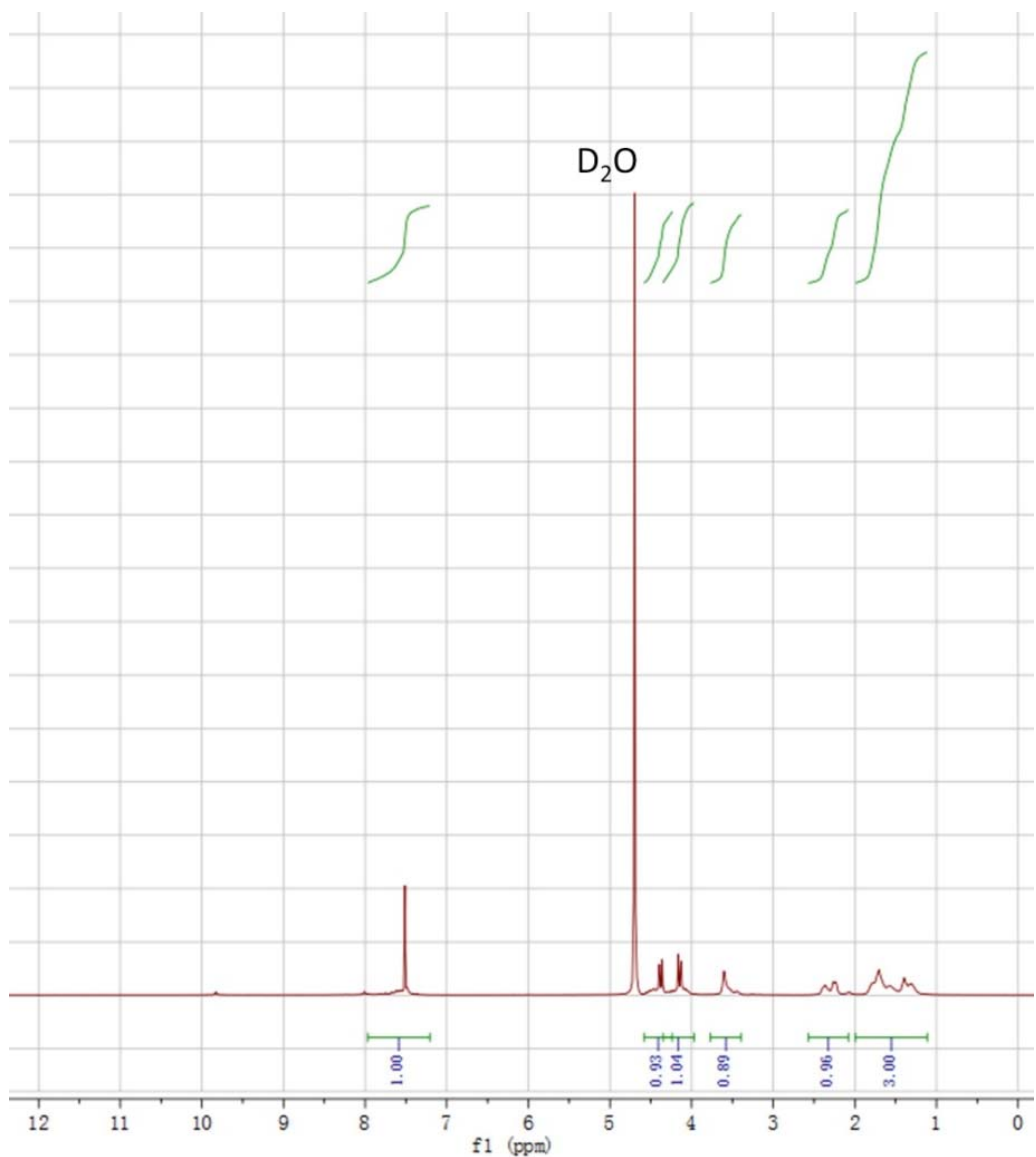

**Figure S38**. The integration of the $^1$H NMR spectrum of the I-Cage-Cl in $D_2O$.

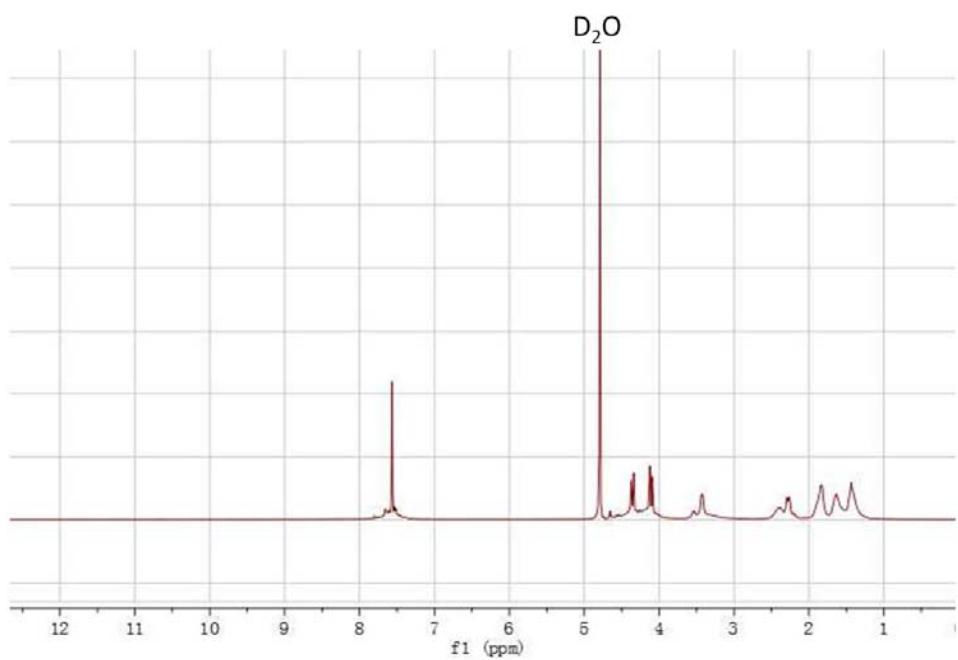

**Figure S39**. The $^1$H NMR spectrum of the Pd@ I-Cage-Cl in D$_2$O.

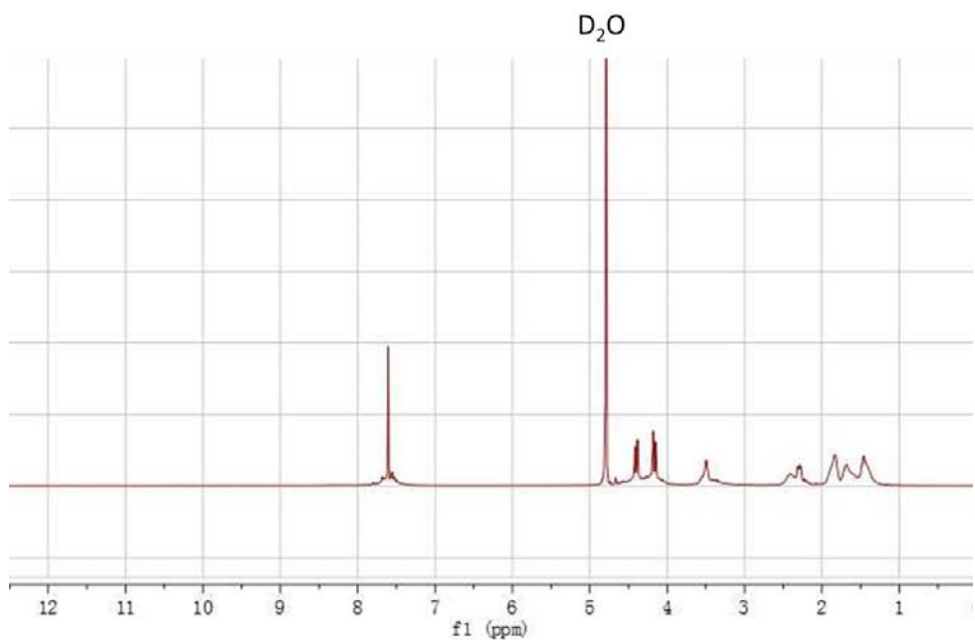

**Figure S40**. The $^1$H NMR spectrum of the Pt@ I-Cage-Cl in D$_2$O.

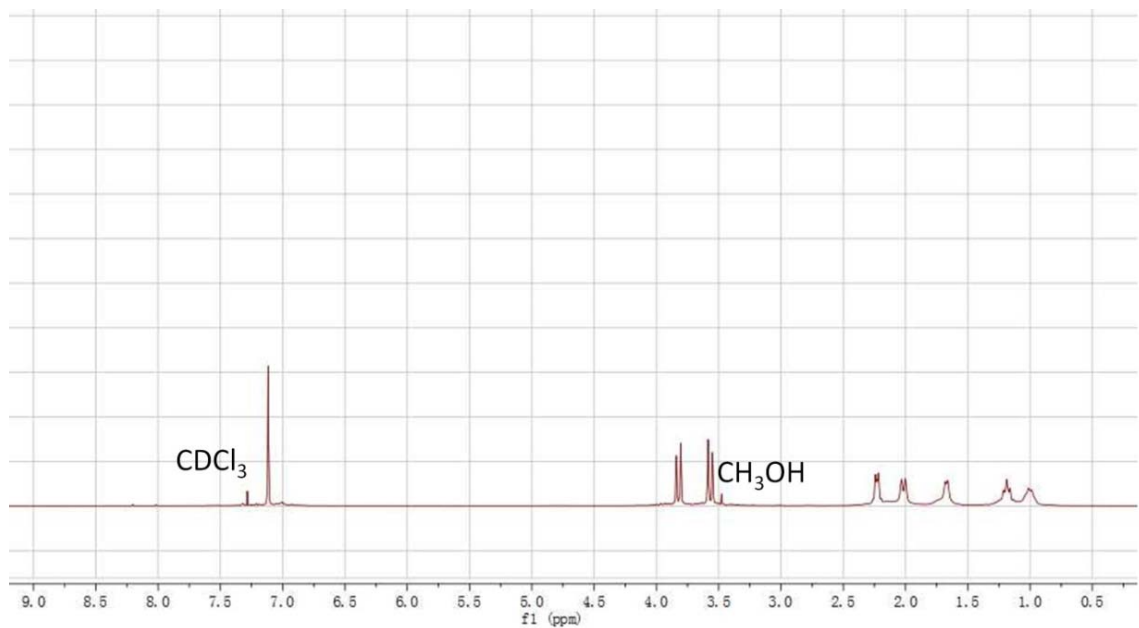

**Figure S41**. The $^1$H NMR spectrum of the Au/RCC3 in CDCl$_3$.